# Superlubric sliding ferroelectricity

Zihao Yang and Menghao Wu

School of Physics and School of Chemistry, Huazhong University of Science and Technology, Wuhan, Hubei, China

Abstract   Sliding ferroelectricity may emerge in many van der Waals bilayers/multilayers and the low switching barriers render ultrafast data writing with low energy cost. We note that such barriers are still much higher compared with structural superlubricity, and in this paper we propose a type of superlubric sliding ferroelectricity in homobilayers separated by a different layer that leads to unprecedented low switching barriers due to incommensurate interfaces. For example, the switching barrier of 3R bilayer $MoS_2$ will be respectively reduced by around 2 or 1 order of magnitudes if they are separated by a graphene or BN monolayer, and the required voltage for switching can be about 1 order of magnitude lower. Such superlubric sliding ferroelectricity widely exists in various similar sandwich trilayer systems where the polarizations stem from symmetry breaking in across-layer stacking configurations, and with ultralow barriers of superlubric sliding, their performances for various applications are greatly enhanced compared with homobilayer sliding ferroelectrics.

Ferroelectrics with switchable electrical polarizations are widely used for various applications. Their required voltage for switching as well as the switching speed depend on switching barriers, which are usually above 100 meV for typical perovskite ferroelectrics. In 2017, we proposed that most 2D materials can become ferroelectric via bilayer or multilayer stacking in either parallel or antiparallel configurations, and the vertical polarizations can be reversed via interlayer sliding upon an electric field.[1] Such so-called sliding ferroelectricity possess unprecedented low switching barriers down to meV (2 orders of magnitude lower compared with most prevalent ferroelectrics), rendering high-speed fatigueless data writing with low energy cost while their room-temperature stability can still be ensured.[2] In later years, its existence has been experimentally confirmed in a series of van der Waals bilayers/multilayers including BN,[3-6] various transition-metal dichalcogenides (TMDs) [7-19], InSe[20,21] and even organic crystals[22], mostly robust at ambient conditions. The high speed, fatigue-resistant and low-voltage data writing has also been realized in devices based on sliding ferroelectrics in recent reports.[23,24]

However, their switching barriers are still much higher compared with superlubric systems with van der Waals incommensurate contacts.[25-34] In such misaligned incommensurate interfaces, frictions almost vanish due to the offset of the atomic lateral force. In previous studies, ultra-low friction coefficients have been demonstrated in heterogeneous bilayers[35-39] (e.g., graphene/$MoS_2$, BN/$MoS_2$, graphene/$SnS_2$, etc.) and also in twisted homobilayers[40,41], which inspires us to consider the possibility of further reducing the switching barriers of van der Waals ferroelectrics via incommensurate contacts. Although superlubricity may exist in homobilayer with a small twist angle that also leads to a moire superlattice of ferroelectric domains,[1,2] the net polarization is compensated so the sliding is not likely to be electrically drivable.

In this paper, we show first-principles evidence that when a sliding ferroelectric homobilayer is separated by a different monolayer with distinct lattice that gives rise to incommensurate contacts, the vertical ferroelectricity may still be formed by the across-layer stacking configuration, which are switchable via superlubric sliding with its unique ultralow barrier. It is noteworthy that polarizations stem from across-layer stacking configurations have been predicted in graphene-based multilayers recently.[42,43] As shown in Fig. 1, for typical sliding ferroelectric switching in homobilayer, the potential energy surface of the commensurate interface leads to a relatively large energy barrier to be overcome in sliding. When such homobilayer is separated by a different layer with distinct lattice, or a homogeneous layer upon a twist angle that also makes the interface incommensurate, the potential energy surface flattens and the barrier can be greatly reduced by such structural lubricity. Meanwhile the vertical polarization may also be alleviated upon the separation of the middle layer. If our aim is to reduce the electric field required for switching, the key point is that the reduction in polarization is trivial compared with the reduction in barrier.

Below we performed density-functional-theory (DFT) calculations implemented in Vienna Ab initio Simulation Package (VASP 5.4) code[44,45]. Projector-augmented wave (PAW) method[46] was used to describe the interaction between valence electrons and atomic cores, while the exchange-correlation potential was treated in the Perdew-Burke-Ernzerhof (PBE)[47] form under generalized gradient approximation (GGA)[48]. A sufficiently large vacuum layer of 40 Å was set in the vertical direction, and the DFT-D2 method of Grimme[49] was used to take van der Waals interaction into account. For structural optimization, the convergence criteria for energy and force are set to be $10^{-6}$ eV and 0.005eV/Å, respectively. The integration was approximated in the Brillouin zone using the Monkhorst-Pack method[50] with a Γ-centered *k*-point grid of 3×3×1, and the plane wave cutoff was set to be 650 eV. The ferroelectric polarizations and switching pathways are obtained respectively by using dipole correction method[51] and nudged elastic band (NEB) method[52].

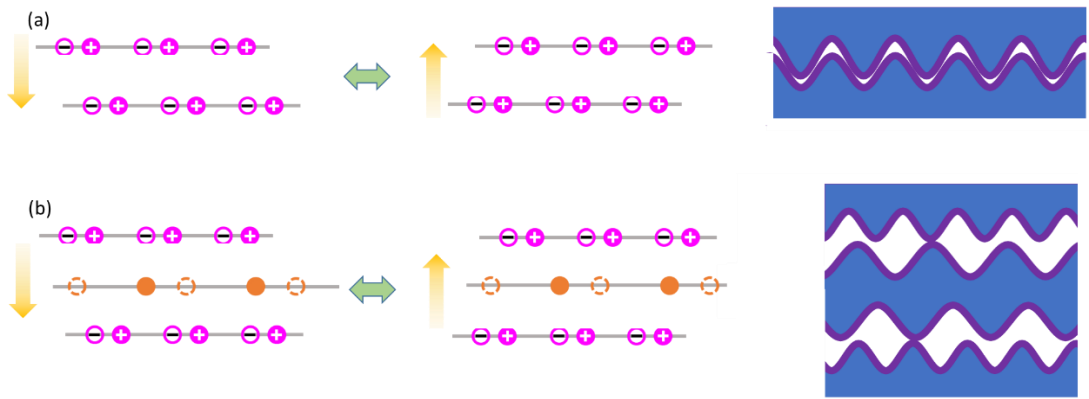

FIG.1 (a) Typical model of sliding ferroelectric switching in homobilayer via interlayer sliding.

(b) Model of ferroelectric switching via superlubric sliding upon insertion of a different monolayer with distinct lattice, where the barriers can be greatly reduced by incommensurate interfaces.

As a paradigmatic case, we take the sliding ferroelectricity in 3R MoS$_2$ as an example. The symmetry breaking in parallel stacking bilayer gives rise to a vertical polarization of 0.5 pC/m, which can be switched via interlayer sliding and the barrier is estimated to be 2.38 meV/atom, as shown in Fig. 2(a), higher compared with bilayer graphene (Gr) or boron nitride (BN). When a graphene or BN monolayer with lattice constants distinct from MoS$_2$ is placed between bilayer MoS$_2$ the symmetry breaking can still be preserved in the across-layer AB stacking configuration between MoS$_2$ bilayer, as shown in Fig. 2(b). Compared with bilayer MoS$_2$, the vertical polarization is reduced to 0.04 pC/m in MoS$_2$/Gr/MoS$_2$ possible due to the charge screening of semi-metallic graphene, while even slightly enhanced to 0.53 pC/m in MoS$_2$/BN/MoS$_2$. Such polarizations can be reversed via relatively sliding of one layer along the armchair direction (accompanied by the sliding of middle Gr/BN layer), transforming the across-layer configuration from AB to BA stacking. The switching barriers are greatly reduced to 0.0267 and 0.167 meV/atom respectively for MoS$_2$/Gr/MoS$_2$ and MoS$_2$/BN/MoS$_2$, around 2 and 1 orders of magnitude lower compared with 2.38 meV/atom for bilayer MoS$_2$. As a result, the required voltage for switching will be around 1 order of magnitude lower for both systems. According to Hirshfeld charge analysis, the charge difference between the top and bottom layer are respectively 0.0003e and 0.007e per unitcell in MoS$_2$/Gr/MoS$_2$ and MoS$_2$/BN/MoS$_2$, in accord with the magnitude of their polarizations. Herein the charge on Gr and BN layer are respectively 0.3328e and 0.5804e per unitcell, indicating the weaker interlayer interaction for the MoS$_2$/Gr/MoS$_2$ system, which can also be revealed by the larger interlayer distance (3.34 Å) of MoS$_2$/Gr/MoS$_2$ compared with MoS$_2$/BN/MoS$_2$ (3.26 Å). If the top layer is dragged along the zigzag direction with the bottom layer fixed, as shown in Fig. 2(c), the vertical polarization will switch frequently along the pathway with the same ultralow sliding barrier, generating an alternating voltage between the top and bottom layer potentially for energy harvesting as a nanogenerator[1] or other nanoelectromechanical applications.

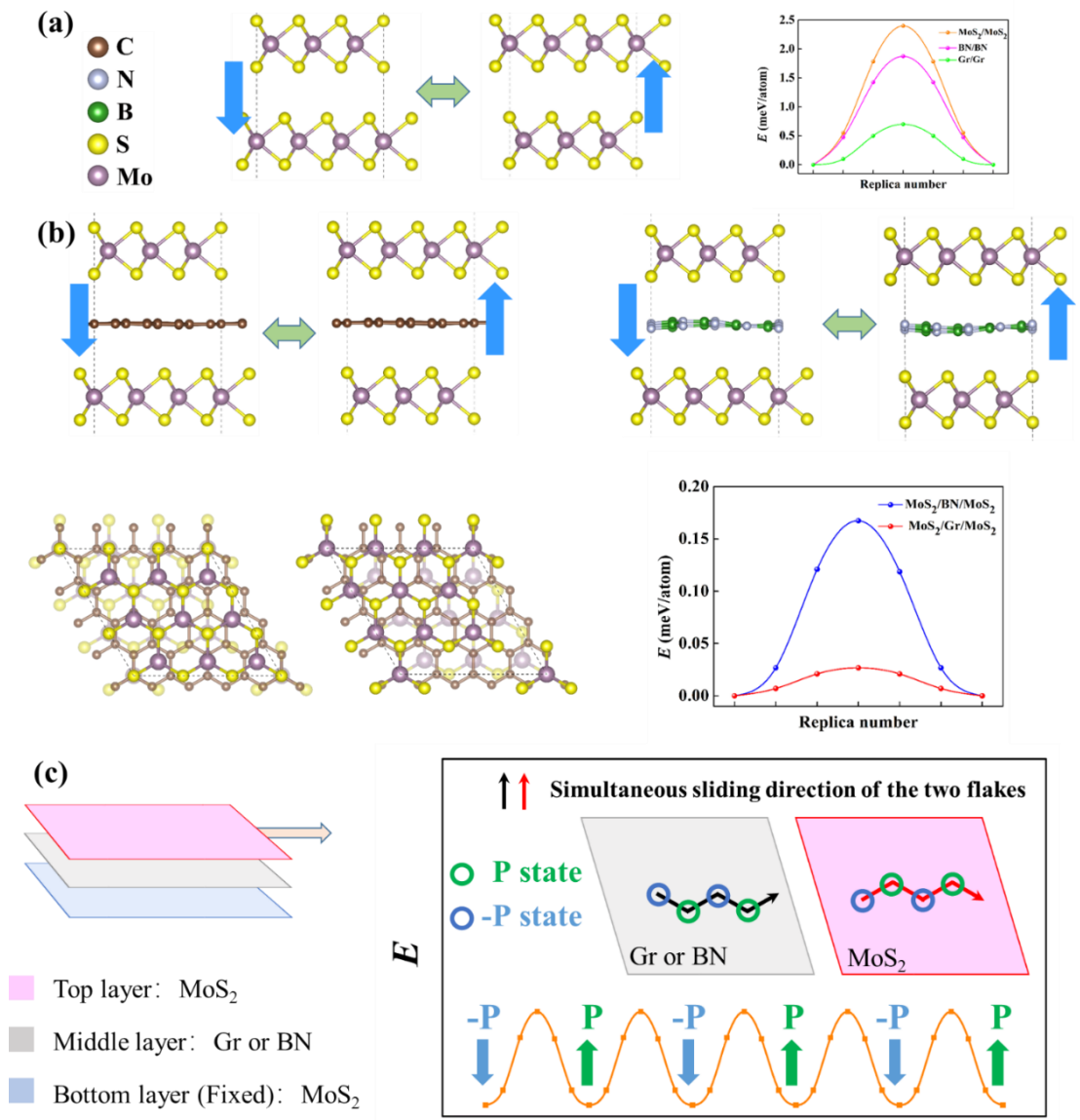

FIG. 2. Ferroelectric switching pathways of (a) bilayer MoS$_2$, (b) MoS$_2$/Gr/MoS$_2$ and MoS$_2$/BN/MoS$_2$ systems, where the blue arrows represent the directions of polarization. (c) The oscillation of vertical polarization along the sliding pathway for MoS$_2$/Gr/MoS$_2$ and MoS$_2$/BN/MoS$_2$ systems, where the top layer is dragged along the zigzag direction with the bottom layer fixed.

The lateral forces are also obtained by the following formula[37,48,53]

$$f_{lateral} = \frac{dE}{dD}$$

where $E$ and $D$ are the energy barrier and corresponding displacement alone the sliding path, respectively. Herein the maximum lateral forces are estimated to be respectively 0.296 and

0.047 meV/Å·atom along the minimum energy sliding path for the MoS$_2$/BN/MoS$_2$ and MoS$_2$/Gr/MoS$_2$ systems, respectively around 2 and 1 orders of magnitude lower compared with bilayer MoS$_2$ (5.16 meV/Å·atom), also revealing high performance compared with 21.79° twisted MoS$_2$/BN heterobilayer(~0.1 meV/Å·atom) reported by Li et al [37].

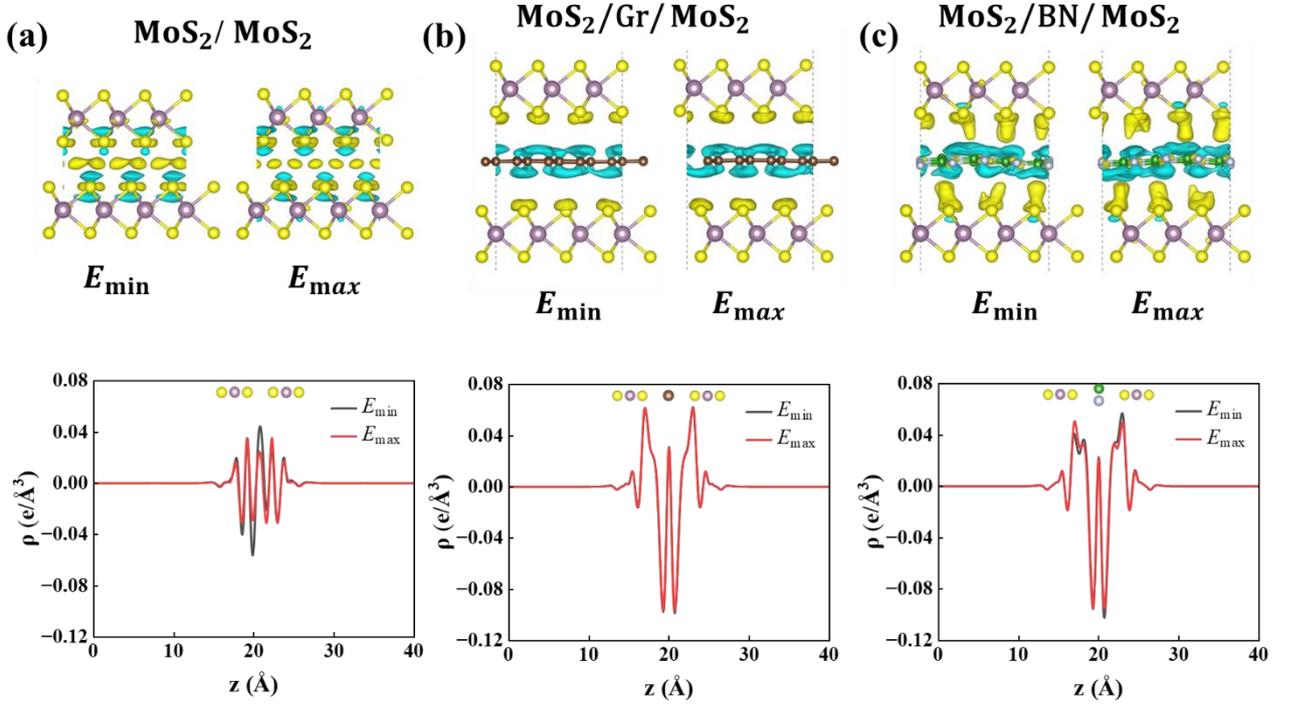

FIG. 3 Differential charge densities at the positions of the energy maxima and minima along the switching pathway and corresponding PDCD along the z direction for (a) MoS$_2$ / MoS$_2$, (b) MoS$_2$/Gr/MoS$_2$, (c) MoS$_2$/BN/MoS$_2$ systems, where the black and red lines correspond to the energy minimum and maximum position, respectively.

To further clarify such superlubricity, the differential charge densities and the corresponding plane-averaged differential charge densities (PDCD) are calculated for the energy maxima and minima locations along the sliding pathway. Essentially, the accumulation and dissipation of electrons in the interfacial region constitutes the fundamental cause of electrostatic interactions between interfaces. When relative motion occurs between two layers at an interface, this resistance caused by electrostatic interaction is to be overcome, and smoother charge density difference fluctuation indicates smaller friction. As shown in Fig. 3, compared with sliding ferroelectric bilayer MoS$_2$ system, the variations of differential charge density for MoS$_2$/BN/MoS$_2$ and MoS$_2$/Gr/MoS$_2$ systems are much reduced during ferroelectric switching, and the smaller variations in electrostatic interaction imply much smaller sliding potential barriers. This also accords with the results of PDCD where their difference between the energy maxima and minima locations are much diminished compared with bilayer $MoS_2$. In particular, the almost overlapping red and black line for the $MoS_2/Gr/MoS_2$ system indicate its high-performance superlubricity, which is also consistent with previous studies on $Gr/MoS_2$ interface[34]. The superlubricity can also be revealed by the variations of interlayer distance during sliding, where the relative difference of interlayer distance between the energy maxima and minima locations $\frac{d_{max}-d_{min}}{d_{min}}$ are respectively 4.1%, 0.37% and 0.12% for homobilayer $MoS_2$, $MoS_2/Gr/MoS_2$ and $MoS_2/BN/MoS_2$ systems.

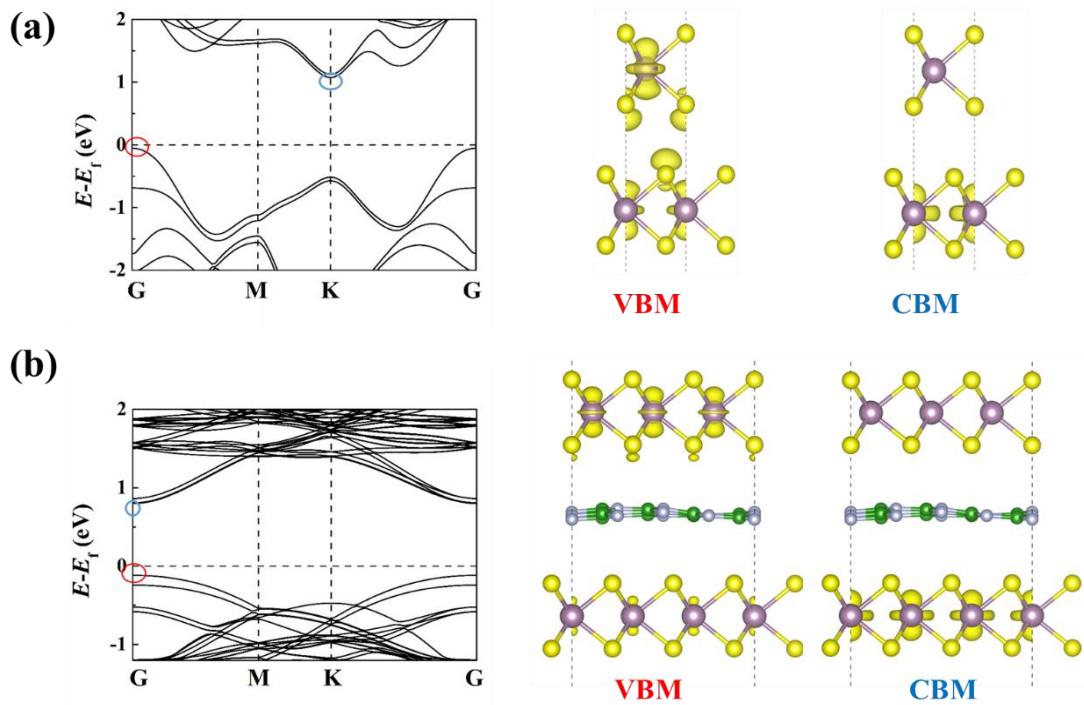

FIG. 4 Bandstructures, VBM and CBM distribution of (a) bilayer $MoS_2$ and (b) $MoS_2/BN/MoS_2$ systems.

The ferroelectric control of exciton is also facilitated by the insertion of BN monolayer in 3R bilayer $MoS_2$. As shown in Fig. 3(a), the bilayer AB stacked $MoS_2$ system is an indirect bandgap semiconductor with 1.13 eV, where the weak vertical polarization induces a small band splitting at the conduction band minimum (CBM) at K point, while the much larger splitting in valence band maximum (VBM) at Gamma point stems from interlayer hybridization.[54] This can be revealed by their distribution in Fig. 4(a), where CBM is mainly distributed on one side while VBM are evenly distributed on both $MoS_2$ layers. However, upon the separation of BN monolayer, the system becomes a semiconductor with a direct bandgap of 0.92 eV in $MoS_2$/BN/$MoS_2$ system in Fig. 4(b). Herein the interlayer hybridization of the $MoS_2$/BN/$MoS_2$ system is much weakened while the vertical polarization is even slightly enhanced, and the band splittings of VBM and CBM at Gamma point are both mainly attributed by the vertical polarization, which can also be revealed by the distribution of VBM and CBM concentrated on different layers. Such separation of electrons and holes as well as the direct bandgap should greatly favor the photovoltaics and photodetection reported in 3R $MoS_2$.[19]

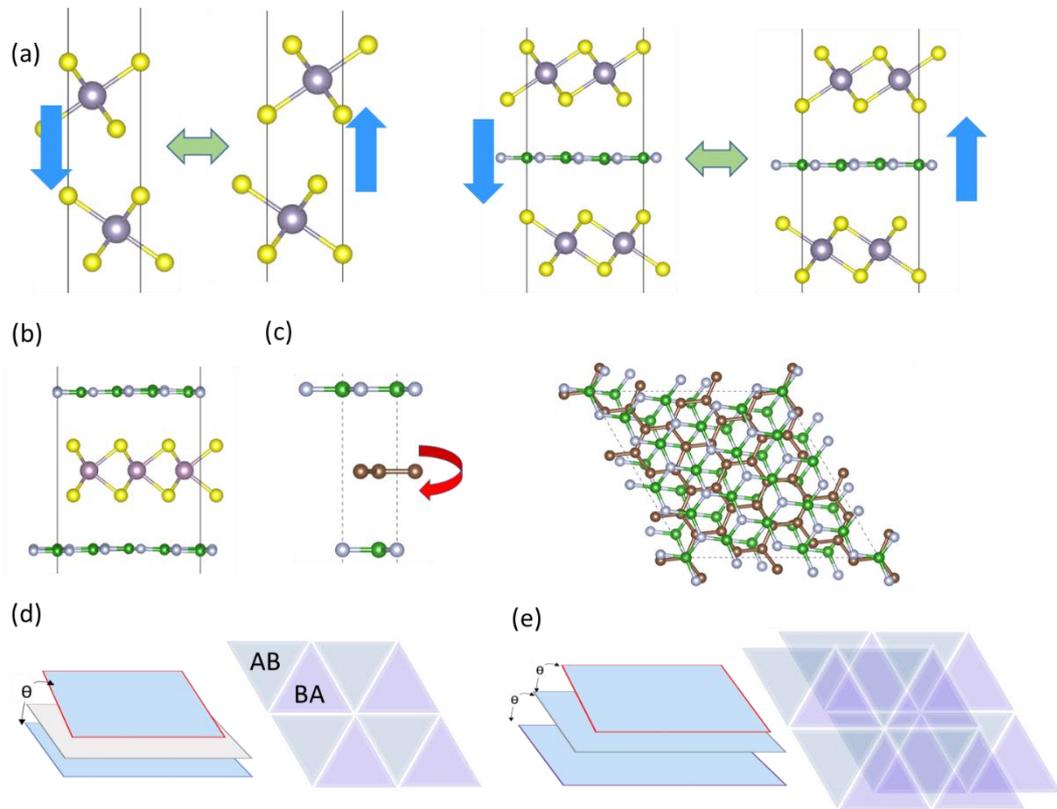

FIG. 5 (a) Ferroelectric switching pathways of (a) bilayer $SnS_2$, and $SnS_2$/BN/$SnS_2$ systems. Geometric structures of (b) BN/$MoS_2$/BN, (c)untwisted and twisted BN/Gr/BN systems. (d) The moire superlattice of homobilayer separated by a different layer, with a small twist angle between the top and bottom layer. (e) The overlapping of two moire superlattices for homotrilayer with small twist angles between adjacent layers, which can be tuned via superlubric interlayer sliding.

The mechanism of superlubric sliding ferroelectricity can be applicable to other similar sandwich systems composed of homobilayers separated by a different monolayer. For example, sliding ferroelectricity may emerge in the antiparallel stacking $SnS_2$ bilayer, while its vertical polarization of 0.16 pC/m remains almost unchanged upon the insertion of BN monolayer. Meanwhile the switching barrier is reduced from 10.7 meV/atom to 0.65 meV/atom, giving rise to superlubric sliding ferroelectricity considering its incommensurate contacts[35] (see Fig. 5(a)), and the required voltage for switching will also be reduced by more than 1 order of magnitude. In some cases the switching barriers as well as the polarizations may diminish simultaneously, e.g., the polarization around 0.016 pC/m for $SnS_2$/Gr/$SnS_2$ is an order of magnitude lower compared with $SnS_2$/BN/$SnS_2$, and the polarization around 0.006 pC/m for BN/$MoS_2$/BN in Fig. 5(b) is more than 2 order of magnitude lower compared with 3R BN bilayer (2.08 pC/m). Upon intercalation of graphene with similar lattice, the untwisted BN/Gr/BN in Fig. 5(c) is 1.95 pC/m, which cannot lead to superlubricity. A large twist angle for the middle layer will make the contact incommensurate for ultralow friction,[39] which simultaneously reduces the barrier and polarization respectively to ~0.01 meV/atom and ~0.02 pC/m at a twist angle of 13. 17 degree. It is noteworthy that if a small twist angle takes place between the top and bottom layer separated by a distinct layer in above systems, ferroelectric moire superlattice with alternating AB (P-up) and BA (P-down) domains (Fig. 5(d)) will be formed akin to homobilayers. If the middle layer is replaced by monolayer of the same material with small twist angles relative to both the top and bottom layer (e.g, twisted trilayer BN or $MoS_2$), the overlapping of two moire patterns can be tuned via interlayer sliding(Fig. 5(e)), which may lead to numerous metastable polar states for neuromorphic computing.[55]

In summary, we predict the existence of superlubric sliding ferroelectricity in a series of sandwich trilayers with incommensurate contacts. In some systems like $MoS_2$/Gr/$MoS_2$, the switching barriers and polarization are respectively reduced by around 2 and 1 order of magnitude, while the barrier is reduced by around 1 order of magnitude for $MoS_2$/BN/$MoS_2$ with polarization even slightly enhanced compared with bilayer $MoS_2$. Such superlubric sliding ferroelectricity does not only minimize the switching barrier to unprecedented low value for ultrafast data writing with low energy cost, but also further facilitate applications like nanogenerator and photovoltaics with enhanced performance compared with homobilayer sliding ferroelectrics.

Acknowledgement

We thank Prof. Deli Peng and Prof. Wei Cao for helpful discussion. This work is supported by National Natural Science Foundation of China (Nos. 22073034).